\documentclass[journal=jacsat,manuscript=article]{achemso}

\usepackage[version=3]{mhchem} 
\usepackage{textcomp}
\usepackage{pdfpages}

\author{Changxin~Dong}
\affiliation{Department of Materials Science \& Engineering, Stanford University, Stanford, CA 94305, USA}
\author{Samya~Sen}
\affiliation{Department of Materials Science \& Engineering, Stanford University, Stanford, CA 94305, USA}
\author{Zhennan~Ru}
\affiliation{Department of Materials Science \& Engineering, Stanford University, Stanford, CA 94305, USA}
\author{Athena~Kolli}
\affiliation{Department of Materials Science \& Engineering, Stanford University, Stanford, CA 94305, USA}
\author{Paxton~S.~Appel}
\affiliation{Classical Conversations, Redwood City, CA 94061, USA}
\author{\textcolor{black}{Jonathan Fan}}
\affiliation{\textcolor{black}{Department of Electrical Engineering, Stanford University, Stanford, CA 94305, USA}}
\author{Eric~A.~Appel}
\email{eappel@stanford.edu}
\affiliation{Department of Materials Science \& Engineering, Stanford University, Stanford, CA 94305, USA}
\alsoaffiliation{Department of Bioengineering, Stanford University, Stanford, CA 94305, USA}
\alsoaffiliation{Stanford ChEM-H, Stanford University, Stanford, CA 94305, USA}
\alsoaffiliation{Institute for Immunity, Transplantation and Infection, Stanford University School of Medicine, Stanford, CA 94305, USA}
\alsoaffiliation{Department of Pediatrics - Endocrinology, Stanford University School of Medicine, Stanford, CA 94305, USA}
\alsoaffiliation{Woods Institute for the Environment, Stanford University, Stanford CA 94305, USA}

\title[Flame retardant aerogels]{Hydrogel-to-Aerogel Transitions in Polymer-Particle Hydrogels Expand the Wildfire Defense Window}

\abbreviations{FTIR, TGA, SEM}

\begin{document}








\pagebreak
\begin{abstract}
    The 2025 Los Angeles wildfires caused widespread urban destruction and displacement, and severe economic losses, highlighting the urgent need for better fire retardants. Current fire suppression strategies rely heavily on water, chemical fire retardants, and water-enhancing gels, which use superabsorbent polymers to retain water and adhere to substrates, offering extended fire protection compared to water alone. However, their effectiveness is limited by evaporation and degradation under extreme heat and wind conditions. This study investigates the thermal properties, evaporation dynamics, and fire retardancy mechanisms of a novel polymer-particle (PP) hydrogel with aerogel-forming capabilities. The boiling-induced water vapor expansion and bubble nucleation drive the transformation of the hydrogel into a highly porous, foam-like fire-retardant coating upon rapid heat desiccation, enhancing thermal insulation. By evaluating the retardancy window across different evaporation stages under high heat and wind conditions, this study aims to determine the duration, effectiveness, and governing physical mechanisms of this unique retardant system. These findings provide a framework for designing the next generation of fire retardants with optimized thermal stability and extended protection for wildfire mitigation.

    \flushleft
    \textcolor{black}{\textbf{Keywords:} Silica aerogel, dynamic hydrogel, fire retardant, mesoporous, thermal activation}
    
\end{abstract}



\section{Introduction}
Wildfires pose an increasing threat to urban environments, causing widespread destruction to homes, critical infrastructure, and local economies. The severity of this impact was starkly visible during the Los Angeles wildfires in early 2025, which resulted in up to \$164 billion in property and capital losses, potentially making it the costliest disaster in U.S. history.\cite{UCLAWildfires2025} Notably, the Palisades and Eaton fires collectively destroyed over 16,000 structures and damaged over 2,000 more.\cite{CALFIRE2025} Beyond structural losses, wildfires disrupt transportation networks and environmental management systems, thereby compounding economic and societal burdens. The expansion of urban areas into fire-prone regions, combined with climate change, has intensified wildfire devastation at the wildland-urban interface (WUI), underscoring the urgent need for more effective fire-retardant technologies and strategies. Yet, existing fire suppression methods that use either water or foams remain limited in their capacity to prevent fire spread and protect critical infrastructure under extreme conditions.

Water-enhancing gels (WEGs) have emerged as a promising class of sprayable fire retardants. These materials are engineered to improve water adherence and extend water retention on surfaces by incorporating common superabsorbent polymers.\cite{USFSWaterEnhancers2025, USFSQPL2020} By forming a hydrated barrier, WEGs enhance fire resistance and suppression effectiveness through delayed evaporation and reduced heat transfer. \textcolor{black}{Unfortunately, conventional WEGs rapidly dehydrate under high-temperature and high-wind conditions common during extreme wildfires, significantly curtailing their protective duration in these environments on account of their water content-dependent retardant efficacy. Studies on the effects of weathering on dehydration and fire performance have underscored the transient nature of these gels, emphasizing the need for formulations that can maintain performance for extended periods in high-risk scenarios.\cite{IBHSGelReport2019} Although the hydration of WEGs functions as a protective heat sink \emph{via} endothermic evaporation, the resultant fire-retardant effectiveness is inherently short-lived due to inevitable dehydration. Moreover, we believe additional improvements are required to fundamentally change the physical structure and mechanical properties of WEGs, such as surface adherence and spreading properties post application, in addition to water retention, for increasing retardant effectiveness.}

Recent investigations into intumescent polymers for fire-protective coatings have inspired a new class of hydrogels, which harness polymer-based fire suppression mechanisms to counter climate change-driven wildfire intensification. Intumescent and aerogel-forming hydrogels represent a significant advancement over conventional intumescent chars, offering improved performance through the formation of protective barrier layers.\cite{Dang2024_AdvMater, Wang_APPPERMEL_2025, Intumescence_Camino_1985, Intumescence_Lim_2016, Intumescence_Guo_2023} Traditional intumescent coatings \textcolor{black}{include epoxy, acrylic, polyvinyl acetate-based paints}, which typically incorporate ammonium polyphosphate, pentaerythritol, and melamine, which react upon heat exposure to form a char layer that insulates the underlying substrate.\cite{andersson2007evaluation, vakhitova2024improving, kalafat2020effect} However, these chars often experience significant shrinkage during combustion, leading to cracking and structural instability that compromise their fire resistance.\cite{ai2019synergistic, chen2021enhancements} Studies indicate that such chars can shrink by more than 50\% during their secondary development, resulting in crack formation and a diminished insulating capability.\cite{song2021evolution} Moreover, while hydrogels offer promise for infrastructure protection, they are typically composed of organic components that inherently exhibit low thermal stability; once desiccated, conventional WEGs lose their effectiveness, limiting their long-term fire protection abilities.

To overcome these limitations, a novel class of water-enhancing hydrogels has been developed that incorporates a hydrogel-to-aerogel transition mechanism to improve thermal stability and flame retardancy.\cite{Dong2024} This innovative approach leverages rapid aerogel formation and inorganic particle densification to transform the hydrogel into a solid aerogel dome, thereby providing enhanced mechanical strength and thermal insulation. These polymer-particle (PP) hydrogels are formed through dynamic multivalent interactions between cellulosic biopolymers and colloidal silica particles (CSP), which promote self-assembly into a viscoelastic hydrogel with robust substrate adherence, shear-thinning properties, and extended fire protection.\cite{Yu2016_PNAS, Yu2019_PNAS, Dong2024} \textcolor{black}{Unlike conventional WEGs, which lose efficacy upon desiccation, the heat-activated aerogel transition in PP hydrogels creates a highly porous, thermally insulating silica coating that effectively prevents substrate ignition. This transformation occurs as rapid water evaporation induces silica particle densification and aerogel formation, thereby maintaining a physical thermal barrier under prolonged flame exposure. The formation of an intumescent aerogel coating and improved mechanical properties, in addition to relaxing the dependence on water content of conventional WEGs, allows the PP hydrogel system to not only delay substrate charring but also demonstrate superior fire resistance compared to commercial alternatives.} These benefits introduce a new paradigm in fire science by offering a longer-lasting, high-performance, and environmentally friendly fire-retardant solution.

Building on our previous research reporting the hydrogel-to-aerogel transitions in PP systems,\cite{Dong2024} we sought to investigate the physical mechanisms underlying PP water-enhancing hydrogel fire protection, analyze its protection window through water loss, and explore the aerogel formation-driven retardancy at polymer-particle interfaces. The findings provide critical insights into expanding the flame retardancy window of intumescent hydrogels, thereby paving the way for the development of next-generation materials capable of more effectively mitigating catastrophic wildfires.

\begin{figure}[!ht]
    \centering
    \includegraphics[width=0.9\textwidth]{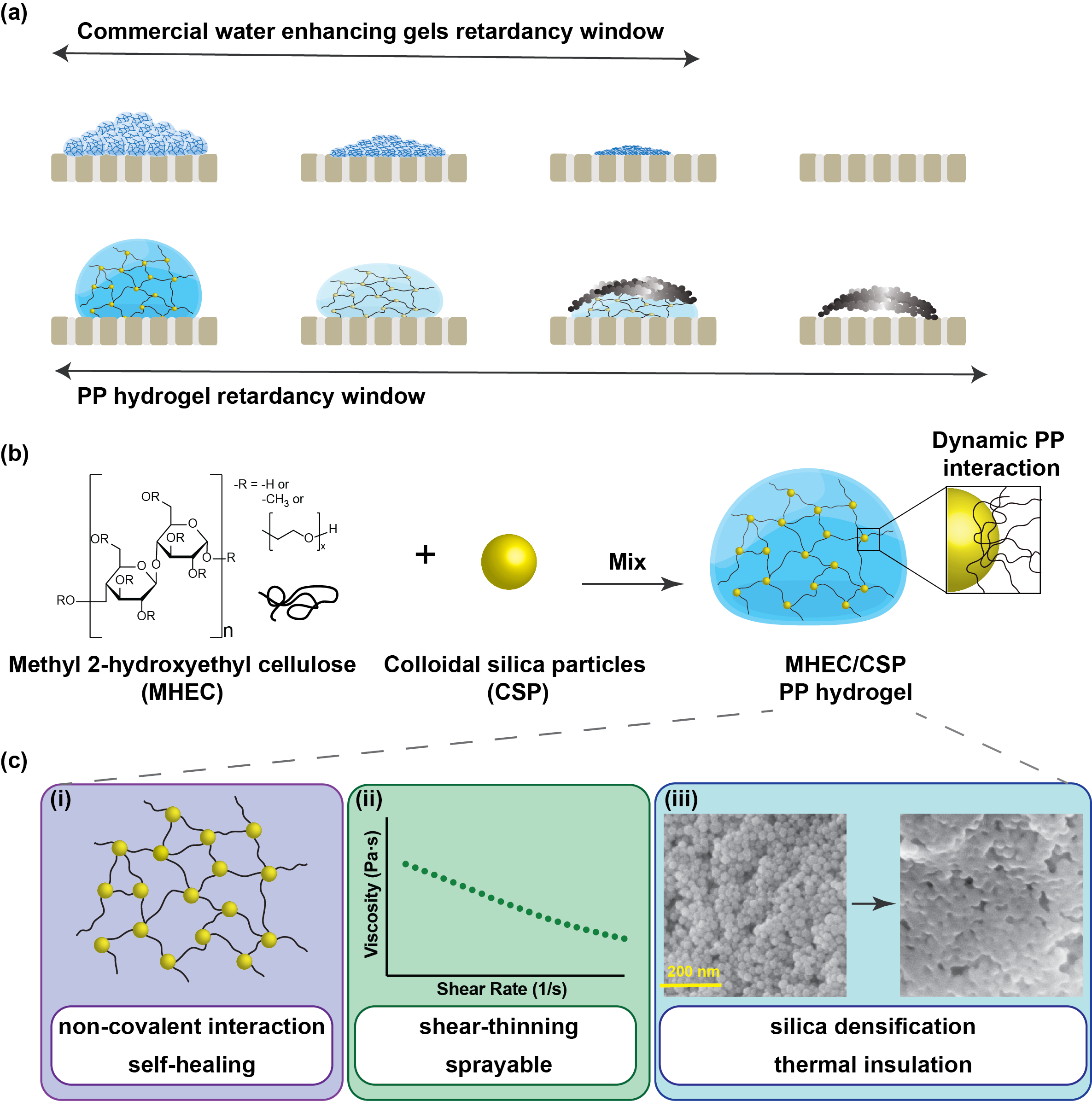}
    \caption{Fire retardancy window and mechanism of water-enhancing gels. (a) Schematic comparing fire protection efficacy of a commercial water-enhancing gel (top) and a polymer-particle (PP) hydrogel, which forms a solid aerogel shell upon heat activation, extending the fire retardancy window during dehydration. (b) Chemical composition of the PP hydrogel, driven by dynamic interactions between methyl-2-hydroxyethyl cellulose (MHEC) and colloidal silica particles (CSP), leading to gelation. (c) Key physical properties contributing to enhanced fire protection: self-healing network structure, shear-thinning behavior for sprayability, and silica densification for improved thermal insulation.}
    \label{fig:fig1_intro}
\end{figure}


\section{Results and Discussion}
In this study we investigate the fire-retardancy window of polymer-particle (PP) hydrogels, focusing on their evaporation dynamics, fire resistance, and foaming behavior. The aerogel effects are examined through the material’s foaming response and microstructural evolution during combustion. To assess performance, we compared a representative PP hydrogel formulation -- methyl hydroxyethyl cellulose crosslinked by colloidal silica particles (MHEC/CSP 1-5) -- against the commercial water-enhancing gel Phos-Chek\textsuperscript{\textregistered} AquaGel-K\textsuperscript{\textregistered} (AquaGK). Unlike AquaGK, which comprises of discrete superabsorbent polymer beads, the PP hydrogel forms a continuous, crosslinked dynamic network that exhibits self-healing behavior (Fig.~\ref{fig:fig1_intro}(a)) \cite{Yu2016_PNAS, Sen2024, Dong2024}. This fundamental difference in microstructure is anticipated to affect both sprayability and fire protection performance.

The selection of MHEC/CSP 1-5 is based on its well-characterized gelation stoichiometry, where colloidal silica particles (LUDOX\textsuperscript{\textregistered} TM-50) dynamically interact with methyl hydroxyethyl cellulose (MHEC) chains through adsorption (Figure~\ref{fig:fig1_intro}(b)) \cite{grosskopf2021gelation, Yu2016_PNAS}. Improved over earlier formulations for industrial manufacturing \cite{Yu2016_PNAS}, this hydrogel exhibits exceptional solution stability due to the minimal conversion of dynamic crosslinks into irreversible covalent bonds, thereby preventing gel aging \cite{Dong2024, Sen2024}. Previous studies have confirmed the gelation mechanism by tracking changes in storage modulus $(G^\prime)$ and loss modulus $(G^{\prime\prime})$ upon the addition of CSPs to MHEC solutions, validating the formation of a robust, biomimetic, and highly scalable hydrogel network that self-assembles through entropically driven polymer-particle interactions \cite{Dong2024, Sen2024}. Furthermore, the commercial availability of both MHEC and CSP ensures practical implementation.

A key advantage of the PP hydrogel over conventional water-enhancing gels, such as AquaGK, is its self-healing property, which maintains structural integrity across the network while preserving sprayability. Rheological analysis \emph{via} dynamic flow sweeps confirms its shear-thinning behavior, a critical attribute for effective spray application; (Fig.~S1) the sprayable nature of PP hydrogels has been verified in an earlier work. Upon exposure to high temperatures, the organic cellulose matrix within the PP hydrogel undergoes controlled charring, leaving behind a densified silicon dioxide (\ce{SiO_2}) network \cite{Dong2024}. This densification is critical for structural integrity, as it solidifies and strengthens the aerogel layer \cite{benad2018mechanical}. In contrast to conventional intumescent chars, which suffer significant shrinkage and cracking during combustion \cite{li2017char, gahane2025exploring}, the PP hydrogel-based aerogel dome remains intact, effectively preserving its thermal barrier properties. The resulting aerogel structure is porous, thermally insulating, and exhibits enhanced structural stability due to silica particle densification \cite{goryunova2024insulating}. This mechanically stable intumescent aerogel represents a significant advancement over traditional fire-protection materials by offering greater durability, extended fire resistance, and improved structural integrity under prolonged high-temperature exposure. This superior performance is demonstrated in Video~S1, where, even after 50\% water evaporation, the PP hydrogel continues to exhibit exceptional thermal insulation and rapid heat-activated foaming, successfully transitioning into an aerogel.


\subsection{Mechanical and material factors governing fire retardancy windows}
Increasing the hydration and altering the thermo-physical property of water-based fire-retardant polymers and gels have been recognized as key factors in improving fire retardancy.\cite{kuznetsov2021physicochemical} However, for aerogel formation-based WEGs, such as the MHEC/CSP aerogels reported by \citeauthor{Dong2024}, the precise mechanisms governing their fire-retardant behavior, retardancy window under evaporation, and intumescence dynamics remain largely unexplored. To address this gap, we systematically examined the behavior of both PP hydrogel and AquaGK systems by preparing formulations at different stages of evaporation (\emph{vide infra}). This experimental approach simulates the progressive water loss over time, allowing us to assess how evaporation influences both fire retardancy and foaming properties.

\begin{figure}[!ht]
    \centering
    \includegraphics[width=0.9\textwidth]{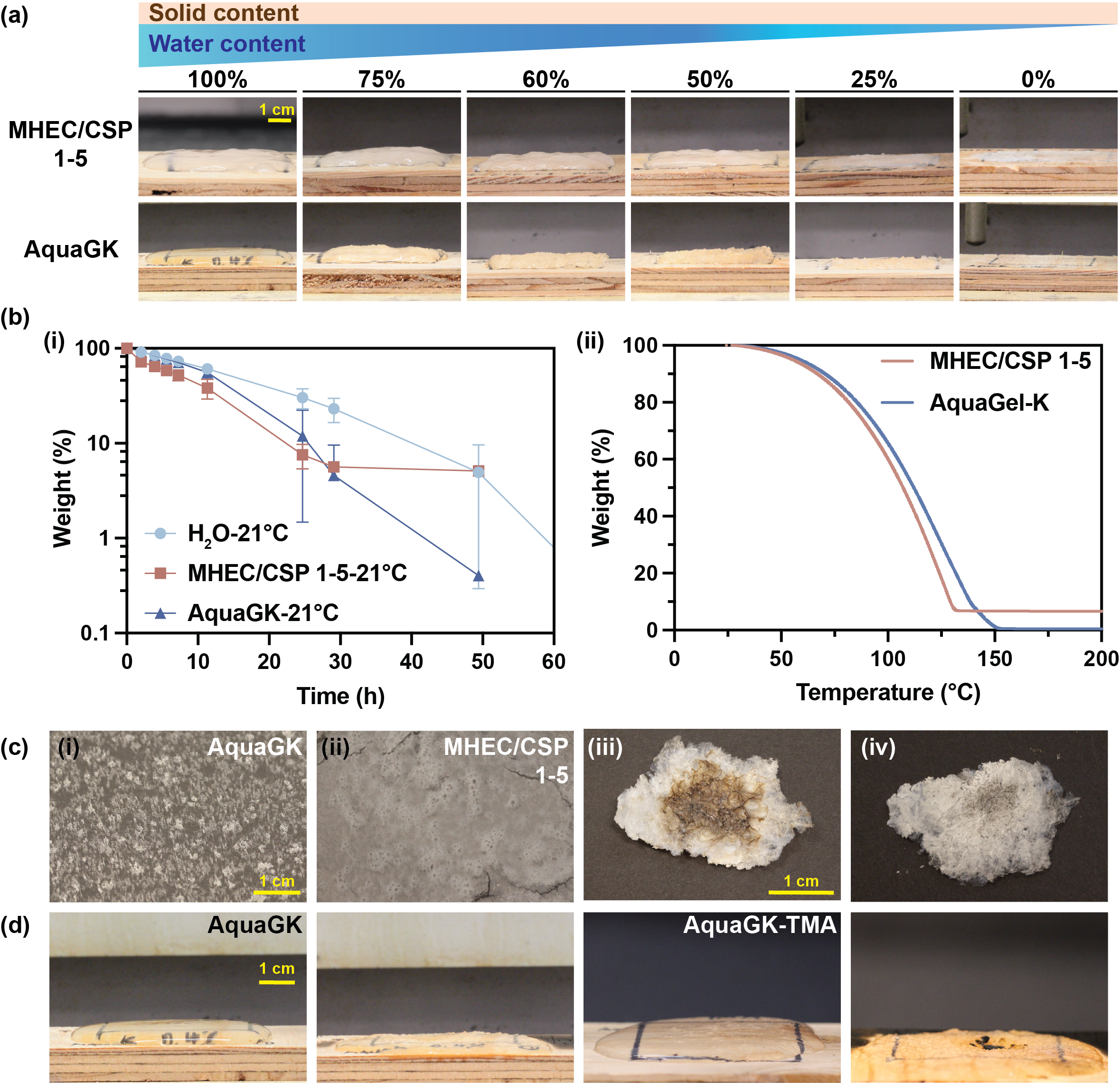}
    \caption{Evaporative properties of MHEC/CSP hydrogels and commercial water-enhancing gels. (a) Sequential images of MHEC/CSP 1-5 and AquaGK during evaporation, showing progressive water loss. (b) Evaporation trends of MHEC/CSP 1-5 and AquaGK under room temperature (i) and thermogravimetric analysis (ii), revealing similar water loss patterns. (c) Desiccated AquaGK (i) and MHEC/CSP 1-5 (ii) post-evaporation, with aerogel formation in MHEC/CSP 1-5 after rapid desiccation via MAP-PRO torch, showing bottom (iii) and top (iv) flame-facing sides. (d) AquaGK and AquaGK-TMA before and after burn tests, demonstrating that silica particle addition to AquaGK does not induce aerogel formation. \textcolor{black}{Error bars in (b-i) indicate standard errors on the mean from triplicate measurements.}}
    \label{fig:fig2_evaporation}
\end{figure}


\subsubsection{Evaporation behavior and its role in fire retardancy}
A widely held assumption in fire suppression materials is that increasing water retention in water-enhancing gels (WEGs) directly improves fire retardancy by prolonging substrate cooling and delaying ignition \cite{IBHSGelReport2019,vahabi2024hydrogel}. However, recent advancements in polymer-particle (PP) hydrogel formulations have challenged this view, demonstrating that hydrogel foaming and subsequent aerogel formation contribute more significantly to fire resistance than water retention alone \cite{Dong2024}. To investigate this hypothesis, we compared the evaporation behavior of the MHEC/CSP 1-5 hydrogel with that of AquaGK under varying conditions (ambient conditions at 21$^\circ$C in open air (Fig.~\ref{fig:fig2_evaporation}(a)), moderate heating at 60$^\circ$C in a controlled thermal chamber (Fig.~S2), and high-temperature exposure at 200$^\circ$C using thermogravimetric analysis (TGA) (Fig.~\ref{fig:fig2_evaporation}(b)(ii))).

Both the MHEC/CSP 1-5 and AquaGK hydrogels exhibited slightly faster evaporation rates than pure water (Fig.~\ref{fig:fig2_evaporation}(b)), suggesting that water retention alone is not the primary determinant of fire retardancy when phenomena such as intumescence, aerogel formation, phase transitions, and physical barrier formation are involved. This observation is consistent with previous studies that indicate the thermophysical properties of water droplets and various fire-extinguishing agents -- including density, heat capacity, and contact angles -- are very similar, resulting in comparable evaporation behaviors \cite{zhdanova2023evaporation}. The minimal differences in evaporation rates between the two hydrogels imply that conventional WEGs, despite being engineered to extend fire protection \emph{via} enhanced moisture retention, may ultimately dehydrate similarly to water under high-temperature wildfire conditions. Notably, even though the MHEC/CSP 1-5 hydrogel exhibited a higher rate of water loss than AquaGK at elevated temperatures, it still demonstrated superior flame resistance (\emph{vide infra}). This reinforces the concept that fire retardancy in PP hydrogels is governed primarily by their heat-activated transition into a thermally insulating aerogel rather than by prolonged water retention alone.


\subsubsection{Gelation mechanism and structural differences}
The gelation mechanisms of MHEC/CSP 1-5 and AquaGK are fundamentally different, influencing their behavior upon desiccation and heat exposure. Upon drying, the films formed by AquaGK (Fig.~\ref{fig:fig2_evaporation}(c)(i)) retain a discrete film pattern, consisting of domains of individually crosslinked superabsorbent polymer beads that remain loosely associated within the dried matrix. In contrast, MHEC/CSP 1-5 hydrogels form a cohesive, uniform polymer network, where colloidal silica particles interconnect cellulose chains, maintaining structural integrity and preventing fragmentation (Fig.~\ref{fig:fig2_evaporation}(c)(ii)). This distinction is crucial, as the continuous network structure of MHEC/CSP 1-5 enables stronger substrate adherence and facilitates its mechanically robust aerogel transformation upon heat activation. This is further demonstrated with the combustion process of AquaGK-TMA in Video~S2, which is not different from AquaGK, Video~S3. Following heat exposure, MHEC/CSP 1-5 undergoes a striking transition into a porous, thermally insulative aerogel (Fig.~\ref{fig:fig2_evaporation}(c)(iii), top view; Fig.~\ref{fig:fig2_evaporation}(c)(iv), cross-section), exhibiting a distinct foamed morphology compared to the thin, non-foaming film observed in AquaGK (Fig.~\ref{fig:fig2_evaporation}(c)(i)). This transformation is driven by rapid foaming aerogel formation and densification of silica particles (Fig.~\ref{fig:fig1_intro}(c)(iii)), enabling the aerogel to retain its structural integrity without collapsing, unlike conventional intumescent chars.


\subsubsection{Role of colloidal silica in hydrogel-to-aerogel transition}
To further investigate the role of colloidal silica particles (CSPs) in the observed hydrogel-to-aerogel transition, we introduced the same 1-5 concentration ratio of alumina-coated silicon dioxide (LUDOX\textsuperscript{\textregistered} TMA) into AquaGK, aiming to induce electrostatic interactions between the negatively charged AquaGK polymeric beads and the positively charged alumina-coated CSPs. This modification was designed to mimic the inorganic-organic interactions found in PP hydrogels and evaluate whether incorporating CSPs into a conventional WEG material could replicate the hydrogel-to-aerogel transition observed in MHEC/CSP 1-5 hydrogels.

Despite the addition of CSPs, AquaGK-TMA formulations failed to exhibit the intumescent foaming behavior characteristic of PP hydrogels (Fig.~\ref{fig:fig2_evaporation}(d), Video~S1, S2). Instead, the material retained a non-foamed, fragmented residue, similar to unmodified AquaGK. This outcome suggests that the mere presence of CSPs is insufficient to drive the hydrogel-to-aerogel transition; rather, the self-assembled polymer-particle network in MHEC/CSP 1-5 plays a critical role in enabling this transformation. The distinct post-combustion behavior underscores a key limitation of AquaGK's discrete hydrogel beads, which lack the network integrity and self-healing properties necessary to support structural expansion during heat activation. \textcolor{black}{Other possible reasons for this behavior could be the limited foaming of the hydrogels due to slow vapor generation, compounded by thermal effects of the aluminum ions, or the fact ther the superabsorbent polymers used in AquaGK don't exhibit surfactant-like behaviors.}


\subsubsection{Hydrogel network integrity and aerogel formation}
The self-healing nature of PP hydrogels is critical in facilitating the hydrogel-to-aerogel transition. When bubbles are introduced as the entrapped water boils upon heat exposure, the PP hydrogels accommodate the commensurate drastic expansion, resulting in the formation of a stable, solidified aerogel structure as the silica densifies and water evaporates. In contrast, the isolated polymeric beads in AquaGK lack the necessary structural connectivity to sustain or stabilize expanding air pockets, thereby preventing the formation of an intumescent layer that exceeds the original hydrated gel volume. These observations underscore the importance of network integrity and dynamic self-healing properties in driving the hydrogel-to-aerogel transformation, which distinguishes PP hydrogels from conventional WEGs and suggests a novel pathway for achieving long-term fire protection in wildfire suppression strategies.

\begin{figure}[!ht]
    \centering
    \includegraphics[width=0.9\textwidth]{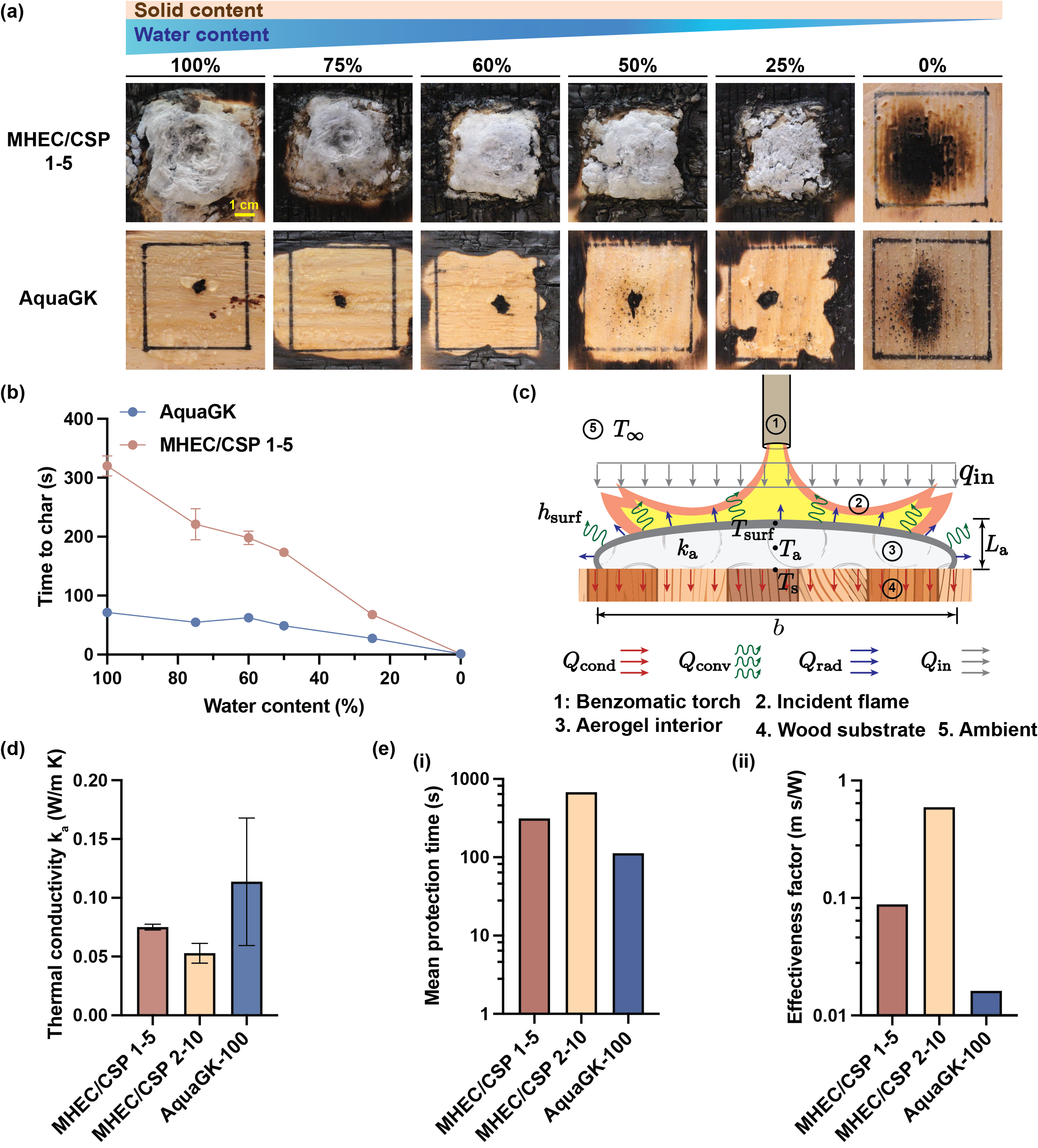}
    \caption{Fire retardancy effectiveness and thermal properties of water-enhancing gels. (a) Post-burn images of MHEC/CSP 1-5 and AquaGK after fire exposure. (b) Time to char as a function of initial water content in MHEC/CSP 1-5 and AquaGK, showing fire retardancy variations during evaporation. (c) Heat-activation model incorporating heat flux, flame interaction, aerogel interior dynamics, wood substrate response, and ambient effects. (d) Thermal conductivity (i), mean protection time (ii), and effectiveness factor (iii) comparisons of MHEC/CSP 1-5, MHEC/CSP 2-10, and AquaGK, highlighting the superior fire retardancy of the stiffer MHEC/CSP 2-10 hydrogel. \textcolor{black}{Error bars in (b, d) indicate standard errors on the mean from triplicate measurements.}}
    \label{fig:fig3_heat-aerogel}
\end{figure}

To further elucidate the role of water content in fire retardancy performance, we evaluated MHEC/CSP hydrogels alongside AquaGK formulations, as shown in Fig.~\ref{fig:fig3_heat-aerogel}. As depicted in Fig.~\ref{fig:fig3_heat-aerogel}(a), fire retardancy does not correlate strongly with water content, given that materials exhibiting different evaporation rates provide similar protective performance. Although the time to char decreases with increasing relative solid content (i.e., lower water content) (cf.~Fig.~\ref{fig:fig3_heat-aerogel}(b)), it is evident that surface coverage and adhesion play more significant roles in maintaining fire resistance. A uniform, well-adhered gel layer effectively prevents direct flame contact and reduces heat penetration, irrespective of its water retention properties, suggesting that prolonging the hydration time alone is not an optimal strategy for enhancing fire retardancy.

Furthermore, while surface adhesion contributes to prolonged protection, our findings indicate that additional thermal mechanisms exert a greater influence on overall fire resistance. Specifically, the material’s ability to modify its thermal conductivity and to form secondary insulating structures -- such as aerogels and silica shells -- upon heat exposure is crucial for enhanced fire protection \cite{Dong2024,Ai_ADMA2025_BoronInSitu,Intumescence_Camino_1985,Intumescence_Lim_2016,Intumescence_Guo_2023}. These structural transformations yield porous, low-conductivity barriers that effectively limit heat transfer, significantly improving insulation and extending the material’s protective lifespan. In contrast, gels that lack these adaptive features depend primarily on evaporative cooling, which is transient and insufficient for long-term fire resistance. Thus, optimizing fire-retardant gels to promote structural changes upon heating appears to be a more effective approach than relying solely on water retention. Achieving this objective requires a deeper understanding of the underlying mechanisms governing thermally activated aerogel formation and the associated heat transfer properties (\emph{vide infra}).


\subsection{\textcolor{black}{Thermal behavior, aerogel formation, and retardant mechanism}}
Thermal conductivity is a directly measurable property of an insulating material and is essential for evaluating the retardancy effectiveness of aerogels. To understand the thermal mechanism behind heat-activated aerogel formation and to calculate the effective thermal conductivity of the aerogels, we model the burn process using the scheme shown in Fig.~\ref{fig:fig3_heat-aerogel}(c). In this model, the total incident heat from the flame torch, $Q_{\rm in}$, is balanced by losses due to radiation, convection, and conduction, such that $Q_{\rm in} = Q_{\rm rad} + Q_{\rm conv} + Q_{\rm cond}$. The input heat is expressed as $Q_{\rm in} = \int\limits_{A} q_{\rm in}\,dS$, where the heat flux from the torch flame is incident on a hydrogel layer of area $A = b^2$. The respective losses are defined as $Q_{\rm rad} = \sigma \epsilon A \left( T_{\rm surf}^4 - T_{\infty}^4 \right), \quad Q_{\rm conv} = h_{\rm surf} A \left(T_{\rm surf} - T_{\infty}\right), \quad Q_{\rm cond} = k_{\rm a} A \left(\frac{T_{\rm surf} - T_{\rm s}}{L_{\rm a}}\right)$. Combining these terms yields the total energy balance as
\begin{align}\label{eq:energy_balance}
    \int\limits_{A} q_{\rm in}\, dS &= \sigma \epsilon A \left( T_{\rm surf}^4 - T_{\infty}^4 \right) + h_{\rm surf} A \left(T_{\rm surf} - T_{\infty}\right) + k_{\rm a} A \left(\frac{T_{\rm surf} - T_{\rm s}}{L_{\rm a}}\right),
\end{align}
which can be solved for the effective mean thermal conductivity of the aerogel, $k_{\rm a}$, using the known experimental parameters (values listed in the SI).

From Fig.~\ref{fig:fig3_heat-aerogel}(d), it is evident that the MHEC/CSP 2-10 hydrogel, which is the stiffest among the formulations, exhibits the lowest effective thermal conductivity, followed by the MHEC/CSP 1-5 formulation, while AquaGK has the highest conductivity. These results correlate with the burn protection performance observed in Fig.~\ref{fig:fig3_heat-aerogel}(a), where MHEC/CSP 1-5 provides more effective burn protection than AquaGK. The lower thermal conductivity associated with MHEC/CSP formulations, resulting from the formation of a robust aerogel layer, leads to a higher protection threshold. Thus, materials with similar water content and water-retention capabilities can exhibit vastly different fire protection and thermal insulation properties, attributable to differences in substrate adherence, coverage, and thermochemical response to heat, particularly via aerogel formation.

As a sanity check, a na\"ive approximation for the effective thermal conductivity of the aerogel can be computed based on the volume fractions of its constituents. Assuming the aerogel is composed of a silica network with air replacing the water in the matrix, for a MHEC/CSP 1-5 hydrogel the volume fractions are $\phi_{\rm silica} \approx 0.05$ and $\phi_{\rm water} \approx \phi_{\rm air} \approx 0.95$. Using the thermal conductivity values for silica $(k_{\rm silica} = 1.3~{\rm W~m}^{-1}{\rm K}^{-1}$) and for air at STP ($k_{\rm air} = 0.025~{\rm W~m}^{-1}{\rm K}^{-1}$), the effective thermal conductivity is defined as $k_{\rm a, eff} \equiv k_{\rm silica} \phi_{\rm silica} + k_{\rm air} \phi_{\rm air}$, yielding $k_{\rm a, eff} \approx 0.089~{\rm W~m}^{-1}{\rm K}^{-1}$. This value is in good agreement with the experimentally calculated $k_{\rm a} = 0.075~{\rm W~m}^{-1}{\rm K}^{-1}$ for MHEC/CSP 1-5 determined from temperature measurements using the model in Eq.~\ref{eq:energy_balance}.

Using the thermal conductivity values shown in Fig.~\ref{fig:fig3_heat-aerogel}(d), we define an ``effectiveness factor'' for hydrogel retardants as $\gamma_{\rm eff} \equiv \frac{\tau_{\rm pr}}{k_{\rm a} \langle T_{\rm s} \rangle}$, where $\langle T_{\rm s} \rangle = \int T_{\rm s}\, dt / \int dt$ is the mean substrate temperature during the burn (essentially a scaled area under the temperature-time curve), and $\tau_{\rm pr}$ is the timescale of protection for the substrate, computed from an exponential fit to the temperature profile $T_{\rm s}(t)$ during the burn. As evident from this definition, a longer protection time, a smaller area under the temperature-time curve, or a smaller effective thermal conductivity results in a higher effectiveness factor. Higher values of $\gamma_{\rm eff}$ indicate a more proficient fire retardant material with superior insulation properties. The values of these quantities for the materials tested in this study are plotted in Fig.~\ref{fig:fig3_heat-aerogel}(e)(i, ii). The data reveal that the mean protection time increases from AquaGK to the MHEC/CSP formulations, with MHEC/CSP 2-10 offering longer protection than MHEC/CSP 1-5. This observation supports the visual evidence from the burn images as well as the thermal conductivity data, and the effectiveness factor distinctly highlights the near-order of magnitude differences among AquaGK, MHEC/CSP 1-5, and MHEC/CSP 2-10 -- the latter exhibiting the highest effectiveness due to its multifaceted function as both a retardant and an insulative material.

This conclusion is further corroborated by thermal imaging of substrate temperatures. In these measurements, the substrate beneath MHEC/CSP 1-5 remains below $100^\circ{\rm C}$, thereby preventing substrate charring by staying within the boiling point of water. In contrast, the substrate beneath AquaGK reaches temperatures exceeding $100^\circ{\rm C}$, resulting in substrate charring due to insufficient thermal insulation (see SI Sec.~3).


\subsection{Hydrogel and aerogel morphology}

\begin{figure}[!ht]
    \centering
    \includegraphics[width=0.9\textwidth]{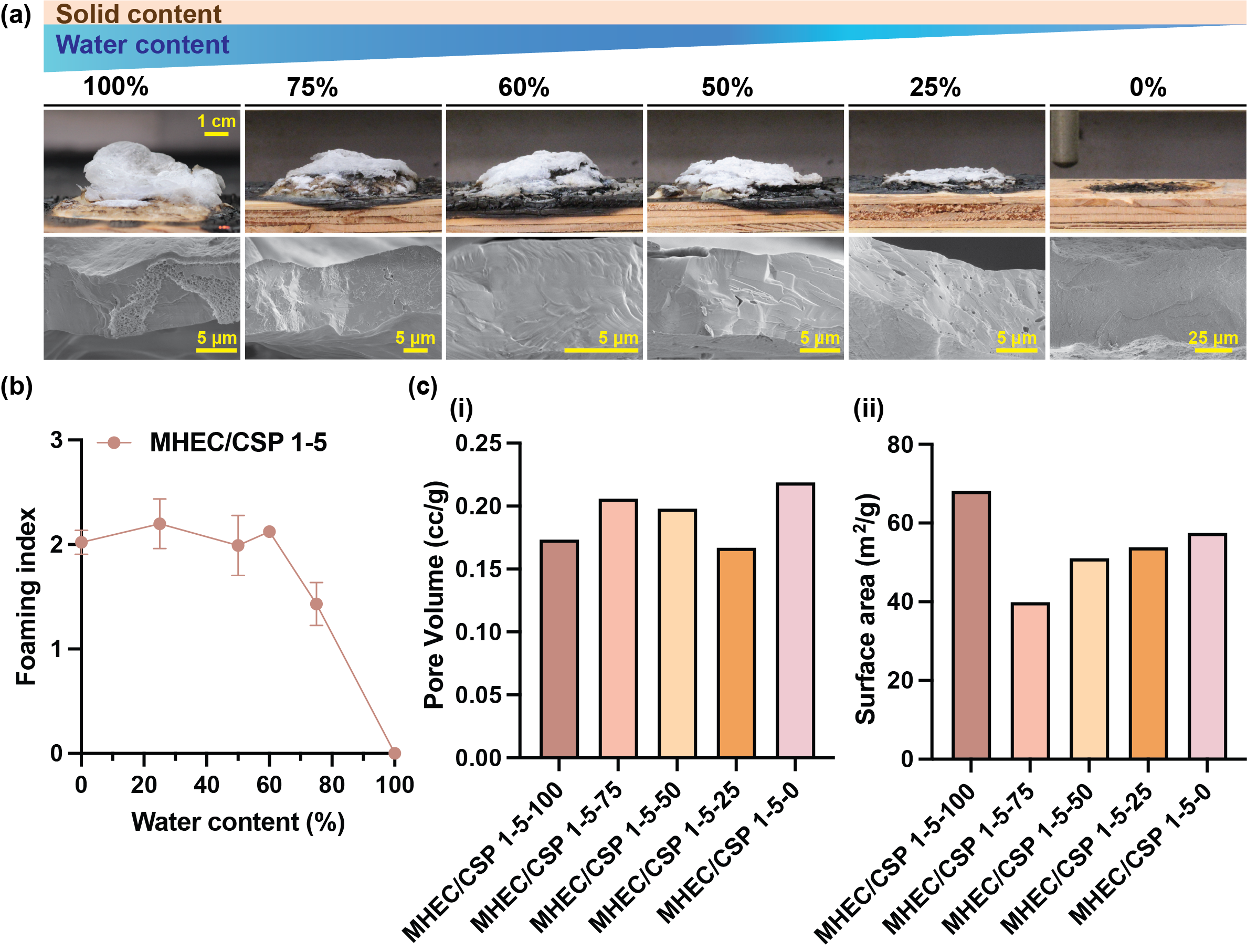}
    \caption{Foaming behavior and porosity evolution of PP hydrogel during evaporation and heat activation. (a) PP aerogel post-heat activation (top) and SEM cross-sections (bottom) showing structural morphology. (b) Foaming index, defined as the aerogel-to-hydrogel thickness ratio, measured across evaporation stages. (c) Pore volume (i) and surface area (ii) of aerogels at different evaporation points, analyzed via Brunauer-Emmett-Teller (BET) measurements. \textcolor{black}{Error bars in (b) indicate standard errors on the mean from triplicate measurements.}}
    \label{fig:fig4_aerogel-morphology}
\end{figure}

\textcolor{black}{MHEC/CSP and AquaGK gels exhibit distinct morphological characteristics and flow behaviors (Video~S4). AquaGK appears as discrete, bead-like hydrogel particles, whereas MHEC/CSP forms a continuous, cohesive hydrogel network. Beyond these visible differences, the MHEC/CSP hydrogel exhibits smoother flow behavior and superior substrate adherence, both of which are crucial for ease of application and enhanced spreadability for surface-coating applications.}
The microstructural evolution of inorganic \textcolor{black}{colloidal silica particles} (CSPs) within the aerogel system has been previously examined using Scanning Electron Microscopy (SEM), Thermogravimetric Analysis (TGA), Fourier Transform Infrared Spectroscopy (FT-IR), and X-ray Photoelectron Spectroscopy (XPS).\cite{Dong2024} However, the role of intumescence-led aerification in micropore formation and overall microstructural evolution remains unexplored. To address this, additional SEM analyses were performed on cross-sections of aerogel films at different stages of evaporation. The results revealed that heat activation, regardless of the evaporation stage, consistently produced aerogel films composed of multiple layers of CSP with a uniform thickness of 5-10~$\mu$m (Fig.~\ref{fig:fig4_aerogel-morphology}(a)). A notable structural difference emerged when the hydrogel was completely dehydrated into a film without heat activation, as shown in the rightmost image of Fig.~\ref{fig:fig4_aerogel-morphology}(a). This dried film exhibited a thickness exceeding 50~$\mu$m, which is 5-10 times thicker than the heat-activated aerogel layers. This observation elicits that rapid exposure to high temperatures is necessary to achieve the characteristic aerogel microstructure, as opposed to passive drying, which results in a bulkier, non-uniform structure.

Despite these microstructural differences, the macroscopic foaming behavior remained largely unaffected by evaporation. The foaming index, defined as the ratio of post-burn aerogel thickness to pre-burn hydrogel thickness, showed no significant decline even after 75\% of the water had evaporated (Fig.~\ref{fig:fig4_aerogel-morphology}(b)). This finding contradicts our initial hypothesis that swelling was integral to the aerogel formation process, instead suggesting that aerogel formation is not hindered by dehydration. Further analysis using Brunauer-Emmett-Teller (BET) surface area measurements confirmed that pore volume and surface area remained consistent across different evaporation stages, with only slight variations. The average pore size remained at 3.4~nm, reinforcing the robustness of the PP hydrogel system in maintaining its flame-retardant properties through substantial water loss that is expected to occur as water evaporates from these materials following application. This pore size classifies the aerogel as a heterogeneous, layered mesoporous structure.\cite{sing1985reporting}

These findings are particularly significant given that the PP hydrogel system does not rely on conventional intumescent mechanisms, such as acid catalysts or gas-blowing agents, yet still achieves effective fire protection. The ability to maintain structural integrity and thermal insulation even after 75\% water loss highlights the resilience and practicality of this hydrogel-based fire-retardant material. The implications of this novel intumescent aerogel formation mechanism, which functions independently of acid and gas-blower chemistry, will be further explored in the next section.


\subsection{Comparison to conventional intumescent char}
It is useful to compare this novel PP hydrogel fire retardant with conventional intumescent char systems, as both enhance fire resistance through volume expansion and thermal insulation mechanisms. Conventional intumescent chars rely on a multi-component chemical reaction triggered at high temperatures, consisting of (i) an acid source or dehydrating agent, (ii) a carbon source or charring agent, and (iii) a blowing agent that generates gas to induce expansion.\cite{gahane2025exploring, davis2015article} While widely used in \textcolor{black}{passive} fire protection \textcolor{black}{applications for buildings, furniture, and industrial infrastructure}, these materials suffer from several critical limitations. One of the primary challenges is the significant shrinkage of the char layer under prolonged heat exposure. This shrinkage and cracking alter the pore size distribution, leading to increased thermal conductivity and reduced fire protection performance.\cite{wang2013thermal} Furthermore, non-continuous and brittle chars are mechanically fragile and prone to delamination, exposing the underlying substrate to direct flame contact and increasing the risk of failure in high-temperature environments. This degradation undermines the char's structural integrity and its ability to provide effective thermal insulation.\cite{gardelle2013thermal} A major limitation of traditional intumescent coatings is the uncontrolled nature of their expansion, as the degree of foaming is highly dependent on the decomposition kinetics of the blowing agent\cite{dreyer2021review} and acid source.\cite{due2021small} Variability in gas release and char formation rates can lead to inconsistent expansion, potentially compromising the coating's protective performance. This inconsistency makes it difficult to achieve a uniform and mechanically stable intumescent layer, limiting the long-term fire protection effectiveness of these materials. These drawbacks highlight the need for more structurally robust fire-retardant materials, particularly for applications requiring extended fire resistance, \textcolor{black}{such as wildfire protection and suppression}.

A key to understanding the aerogel-forming behavior of PP hydrogels is the boiling of the entrapped water in the hydrogel structures, leading to expansion of the aerogel structure. Moreover, the thermal degradation of cellulose ethers releases various volatile products, further driving the expansion of the aerogel structure.\cite{janigovadegradability} These events contribute to the formation of a porous, expanded char structure, thereby enhancing fire-retardant properties. Additionally, gas bubbles introduced during the mixing and spraying process expand upon heating, further enhancing the foaming effect. These gases formed by water boiling and cellulose ether degradation facilitate the formation of a structured aerogel layer by expanding within the hydrogel matrix during heat exposure. The controlled gas release ensures consistent volume expansion and uniform spacing between silica layers, as confirmed by SEM cross-sections (Fig.~\ref{fig:fig4_aerogel-morphology}(a)), ultimately stabilizing the aerogel structure. Furthermore, the sintering of \ce{SiO_2} particles plays a crucial role in maintaining structural integrity. Unlike gas-generating processes, silica sintering occurs \emph{via} heat-driven atomic diffusion, leading to particle coalescence and densification into a continuous solid network. Since \ce{SiO_2} is already in an oxidized state, it does not undergo decomposition or release gaseous byproducts during sintering.

Another key advantage of the PP hydrogel system is its self-healing and viscoelastic properties, which enable controlled foaming and expansion without forming brittle fractures. Unlike conventional chars, which often result in rigid and fragile structures, the dynamic polymer-particle interactions within the hydrogel network allow for a flexible and mechanically resilient fire-retardant layer. This interconnected polymer-silica network not only enhances fire resistance but also significantly improves mechanical durability, ensuring that the protective layer remains effective throughout the fire event.

Thus, unlike conventional intumescent chars, the PP hydrogel-based intumescent system introduces a fundamentally different approach by leveraging a hydrated polymer matrix that provides an initial heat sink through endothermic water evaporation. This hydrogel-based cooling mechanism delays heat penetration and reduces surface temperatures, effectively extending the protection window. The incorporation of inorganic CSP further enhances thermal stability, as the silica undergoes densification upon heating, forming a highly porous and mechanically stable aerogel structure. This transformation prevents the structural collapse commonly observed in conventional char layers, ensuring that the aerogel layer remains intact and maintains its insulating properties even under prolonged flame exposure. The hydrogel-to-aerogel transition represents a fundamental shift from traditional intumescent mechanisms, introducing a new class of fire-retardant materials that integrate multiple protective strategies: hydrogel-based cooling, inorganic aerogel formation, and dynamic polymer-particle interactions. This novel approach offers several key advantages, including extended fire protection duration, enhanced substrate adherence, improved environmental stability, and practical applicability in fire suppression technologies. By overcoming the inherent weaknesses of conventional intumescent coatings, the PP hydrogel system presents a next-generation solution for advanced fire protection in high-risk environments, particularly in applications where long-term thermal insulation and structural integrity are paramount.


\section{Conclusion}
In this study, we demonstrated that a dynamically crosslinked hydrogel comprising robust PP interactions exhibits exceptional foaming ability and fire retardancy, even under dehydration, making it a promising alternative to conventional superabsorbent polymer based water-enhancing hydrogels. Unlike traditional formulations that rely primarily on water retention for fire suppression, the PP hydrogel system leverages an integrated self-healing and intumescent mechanism to achieve superior fire protection. This intumescent process is driven by the controlled degradation of cellulose ether, which releases gaseous byproducts that facilitate expansion, while concurrent silica densification stabilizes the aerogel structure, yielding a mesoporous, uniformly layered thermal barrier.

\textcolor{black}{Notably, these PP hydrogel materials are complementary to conventional intumescent coatings for structure protection. While traditional intumescent coatings may be applied to structures to harden them to wildfire, PP hydrogels function similar to other WEGs and are applied during wildfire scenarios to protect from oncoming fire. These PP hydrogels have exhibit the added benefit of a heat-activated hydrogel-to-aerogel transition that enhances the wildfire defense window, enabling longer-lasting protection against structural damage from wildfire.}

Beyond robust fire-retardant performance, the PP hydrogel system is inherently biodegradable,\cite{Dong2024,Yu2019_PNAS} reducing environmental impact compared to conventional synthetic fire suppressants. The scalability and adaptability of these materials make them a compelling candidate for next-generation fire-retardant platforms, offering a sustainable, high-performance solution for structural protection in catastrophic wildfire scenarios. The findings from this study pave the way for further exploration of hydrogel-based intumescent systems, potentially revolutionizing fire protection strategies for infrastructure, wildland-urban interfaces, and critical assets exposed to extreme fire events.

\section{Materials and Methods}

\subsection{\textit{Materials}}
Phos-Chek AquaGel-K powder was provided by Perimeters Solutions. All other materials were purchased from Sigma-Aldrich and used as received.

\subsection{\textit{Polymer-Particle Hydrogel Preparation}}
Hydrogel formulations were prepared as follows. Polymer solutions were first made by dissolving methyl 2-hydroxyethyl cellulose (MHEC) (viscosity 15,000~cP, Sigma-Aldrich) in water at a concentration of 60~mg/mL. The solution was stirred vigorously at room temperature overnight to ensure complete dissolution. The colloidal silica particles (CSP) used in this study were LUDOX\textsuperscript{\textregistered} TM-50 (Sigma-Aldrich), a dispersion of amorphous anionic silica with a particle size of 22~nm in diameter, produced by polarizing silica nuclei from silicate solutions under alkaline conditions. CSP is commonly used as a frictionizing agent in paper and textile coatings. To achieve uniform dispersion in the gel formulation, CSP was diluted from 50~wt\% to 15~wt\% (pH~9) with water and then slowly added drop-wise into the MHEC solution while stirring with a spatula at a 1:5 weight ratio. Additional water was added to reach the final gel concentration.

\subsection{\textit{Evaporation Behavior}}
The evaporation behavior of MHEC/CSP 1-5, AquaGK, and water was monitored at room temperature ($21^\circ{\rm C}$) and 30-35\% humidity by measuring weight loss in a petri dish over 60~h. Samples had an initial weight of 16-20~g.

\subsection{\textit{Thermogravimetric Analysis}}
Thermogravimetric analysis (TGA) was conducted using a TA Discovery TGA 5500 at the Stanford Soft \& Hybrid Materials Facility (SMF). MHEC/CSP 1-5 and AquaGK were placed on a platinum sample pan and observed for weight loss by heating at a rate of $20^\circ{\rm C}/{\rm min}$ from room temperature to $200^\circ{\rm C}$, then holding isothermally at $200^\circ{\rm C}$ for 20~min. 

\subsection{\textcolor{black}{\textit{Wildfire-Relevant Flame Tests}}}
\textcolor{black}{We utilized small-scale burn tests adapted from ASTM E1321-1997(02), as recommended by the U.S. Forest Service’s Wildland Fire Chemicals program. \cite{Dong2024,usfs_wfcs_weg}} A 20~g hydrogel sample was applied to a plywood plate, covering a 5~cm $\times$ 5~cm square area (Lowe's Unfinished Whitewood Board, cut into 10~cm $\times$ 10~cm squares). A Bernzomatic MAP-Pro Premium Hand Torch, fueled by propylene (99.5-100~vol\%) and propane (0-0.5~vol\%), capable of reaching temperatures up to $2054^\circ{\rm C}$, was used for burning tests. The torch was stabilized approximately 5~cm above the hydrogel surface. Burn time was measured from the moment the torch was ignited to the point when the wood substrate visibly began to char.

The foam thickness was determined as the distance between the top of the aerogel foam and the surface of the wood plate. Post-burn, the aerogel foam was collected for SEM imaging and BET analysis. The temperature within the aerogel was recorded using a K-type thermocouple (0 to $1300^\circ{\rm C}$), connected to a 4-channel thermocouple thermometer data logger. The substrate temperature was recorded using a TOPDON TC004 Thermal Imaging Camera within 30 seconds of aerogel removal after the burn.

\subsection{\textit{Scanning Electron Microscopy}}
The morphology and thickness of the burnt aerogel cross-section were characterized using scanning electron microscopy (SEM) with an FEI Magellan 400 XHR at the Stanford Nano Shared Facility (SNSF). Samples were mounted on 90-degree aluminum pin stubs using double-sided conductive copper tape. To minimize charging, a 4~nm layer of pure gold was deposited onto the samples using a Leica ACE600 vacuum system before imaging. Imaging analysis was conducted at a beam voltage of 3~kV under high vacuum in field-free mode.

\subsection{\textit{BET Analysis}}
Pore characterization was conducted using Anton Paar NovaTouch surface area and pore size analyzer at Stanford Soft \& Hybrid Materials Facility (SMF). Approximately 80~mg of the powder are placed into a pre-weighed glass tube and sealed with a rubber ring. The tube is vacuumed and dried under $80^\circ{\rm C}$ for 1~h, $120^\circ{\rm C}$ for 1~h, and $150^\circ{\rm C}$ for 6~h to remove any moisture. After drying, the evacuated cell is weighed, and the initial tube weight is deducted to ascertain the precise mass. The sample was characterized from \ce{N_2} adsorption/desorption isotherm process at 77~K. Surface area was calculated using Brunauer–Emmett–Teller (BET) method based on the \ce{N_2} adsorption process with BET relative pressure ($P/P_0$) range of 0.05 to 0.3. Pore volume and pore size distribution were calculated using the Barrett-Joyner-Halenda (BJH) method based on the \ce{N_2} desorption data.

\subsection{\textit{Rheological Characterization}}
The rheological properties of the hydrogels were measured using a 20~mm serrated parallel plate on a stress-controlled TA Instruments Discovery HR-2 rheometer at a gap of $500~\mu{\rm m}$. Flow sweeps were performed at shear rates of $10^2$ to $10^{-3}~{\rm s}^{-1}$ with steady-state detection.


\begin{acknowledgement}
The authors appreciate the Gordon \& Betty Moore Foundation for their financial support of this work as part of our efforts to develop wildland fire retardants to improve the execution of prescribed burns and enhance forest management (Grant Numbers 11539). Part of this work was performed at the Stanford Nano Shared Facilities (SNSF), supported by the National Science Foundation under award ECCS-2026822. Part of the rheology data was collected at the Stanford Soft and Hybrid Materials Facility (SMF), supported by the NSF grant National Science Foundation grant ECCS1542152. \textcolor{black}{The authors gratefully acknowledge the support of the Stanford Strategic Energy Research Consortium (Award No. 327778) for Brunauer–Emmett–Teller (BET) surface area measurements.}
\end{acknowledgement}

\section*{Conflicts of Interest}
E.A.A. is an inventor on a patent that describes the technology reported in this manuscript. All other authors declare no conflicts of interest.

\begin{suppinfo}
Data on the flow rheology of hydrogels, evaporation properties, thermal imaging of burn tests, and conductivity modeling are in the Supplemental Information. Burn tests are shown in the Supplementary Videos S1-S3. \textcolor{black}{Video S4 shows the difference in flow properties of AquaGK and MHEC/CSP hydrogels, with water as the control.}
\end{suppinfo}


\bibliography{references}

\includepdf[pages=-]{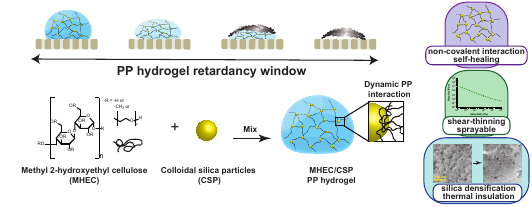}

\end{document}